\documentclass[prl,twocolumn,nofootinbib,nobibnotes,noeprint,preprintnumbers,floatfix]{revtex4-2}
\usepackage{amssymb,amsmath,graphicx}

\usepackage[utf8]{inputenc}
\usepackage{amsmath,amssymb}
\usepackage{hyperref}
\usepackage{xcolor}
\usepackage{mathtools}

\def\lsim{\mathrel{\raise.3ex\hbox{$<$\kern-.75em\lower1ex\hbox{$\sim$}}}}
\def\gsim{\mathrel{\raise.3ex\hbox{$>$\kern-.75em\lower1ex\hbox{$\sim$}}}}

\begin{document}

\preprint{FERMILAB-PUB-23-291-T}

\title{Dark Matter Is The New BBN}
\author{Dan Hooper$^{1,2,3}$}
\author{Huangyu Xiao$^1$}
\affiliation{$^1$Fermilab, Theoretical Astrophysics Department, Batavia, IL 60510, USA}
\affiliation{$^2$University of Chicago, Kavli Institute for Cosmological Physics, Chicago IL, USA}
\affiliation{$^3$University of Chicago, Department of Astronomy \& Astrophysics, Chicago IL, USA}
\date{\today}

\begin{abstract}

Measurements of the primordial element abundances provide us with an important probe of our universe's early thermal history, allowing us to constrain the expansion rate and composition of our universe as early as $\sim 1 \, {\rm s}$ after the Big Bang. Prior to this time, we have essentially no empirical information on which to base any such claims. In this paper, we imagine a future time in which we have not only detected the particles that make up the dark matter, but have measured their mass and annihilation cross section with reasonable precision. In analogy to the light element abundances, the dark matter abundance in this scenario could be used to study and constrain the expansion rate and composition of our universe at the time of dark matter freeze out, which for a standard thermal relic occurs at $T_f \sim m_{\chi}/20$, corresponding to $t \sim 4 \times 10^{-10} \, {\rm s} \times ({\rm TeV}/m_{\chi})^2$, many orders of magnitude prior to the onset of Big Bang nucleosynthesis. As examples, we consider how such measurements could be used to constrain scenarios which feature exotic forms of radiation or matter, a ultralight scalar, or modifications to gravity, each of which have the potential to be much more powerfully probed with dark matter than with the light element abundances.

\end{abstract}
\maketitle
\section{Introduction}

Observations of the cosmic microwave background, large scale structure, and the rate of Hubble expansion have provided us with a detailed description of our universe's history, spanning from the time of recombination to the present. To study and constrain our universe prior to recombination, we rely almost entirely on  measurements of the primordial element abundances~\cite{Cooke:2013cba,Riemer-Sorensen:2017vxj,Izotov:2014fga,Aver:2015iza,Peimbert:2016bdg}. These measurements, which are in good agreement with the predictions of Big Bang nucleosynthesis (BBN), provide us with our earliest glimpse into our universe's thermal history (for a summary, see the Particle Data Group's review of BBN~\cite{ParticleDataGroup:2022pth}). In particular, this data confirms that our universe was radiation dominated and generally well described by the standard $\Lambda$CDM cosmological model during and after the era of primordial nucleosynthesis~\cite{Schramm:1997vs,Steigman:2007xt,Iocco:2008va,Cyburt:2015mya,Pitrou:2018cgg,Fields:2019pfx}. Furthermore, these measurements allow us to significantly constrain the composition of our universe and the laws that governed its evolution as early as the time of proton-neutron freeze out, $\sim 1 \, {\rm s}$ after the Big Bang, when the temperature was $\sim 1 \, {\rm MeV}$~\cite{Burns:2022hkq,Depta:2019lbe,Kawasaki:2017bqm,Forestell:2018txr,Hufnagel:2017dgo,Hufnagel:2018bjp,Huang:2017egl,Poulin:2015opa,Pospelov:2010hj,Cyburt:2009pg,Jedamzik:2009uy,Cyburt:2002uv,Sarkar:1995dd}.

Despite everything we have learned from the primordial element abundances, we have essentially no empirical information to constrain the thermal history of our universe prior to BBN. Although we can use the Large Hadron Collider (LHC) and other accelerators to study the interactions of Standard Model particles at energies which correspond to temperatures as high as $\sim 1 \, {\rm TeV}$ ($\sim 10^{-12} \, {\rm s}$), new physics that is not significantly coupled to Standard Model fields remains largely unconstrained by laboratory experiments. It thus remains possible that new physics may have played an important role in the evolution of our early universe prior to the onset of BBN. As a few examples, our early universe may have included an era that was dominated by exotic matter or radiation~\cite{Banks:1993en,deCarlos:1993wie,Kamionkowski:1990ni,Berlin:2016gtr,Berlin:2016vnh,Hooper:2019gtx,BhupalDev:2016gna,Gelmini:2006pq,Kane:2015jia,Randall:2015xza,Nelson:2018via}, or in which the masses and/or couplings of particles were influenced by an evolving scalar field~\cite{Delos:2023vfv,Gan:2023wnp,DiMarco:2018bnw,DEramo:2017gpl,Davoudiasl:2015vba,Gardner:2004in,Choi:1999xn,Joyce:1996cp,Spokoiny:1993kt,Ringwald:2014vqa,Nelson:2011sf, Graham:2015rva}. Unknown phase transitions may have occurred at very early times~\cite{Athron:2023xlk,Boileau:2022ter,Hindmarsh:2020hop,Chung:2011hv,Quiros:1999jp,Kamionkowski:1993fg}, and modifications to general relativity~\cite{Dutta:2016htz,Meehan:2014bya,Clifton:2011jh,Bertschinger:2008zb,Catena:2004ba,Okada:2004nc,Ahmed:2002mj,Stelle:1976gc,Wagoner:1970vr} could have lead to a wide range of exotic dynamics.

In this paper, we imagine a point in the future when the particle species that makes up the dark matter has been identified and its characteristics have been broadly measured. If the dark matter is a thermal relic of the early universe, measurements of its annihilation cross section could be used to estimate the relic abundance that is predicted for this state (for reviews, see Refs.~\cite{Bertone:2004pz,Lisanti:2016jxe,Arcadi:2017kky}). This is in close analogy to how our knowledge of various nuclear reaction rates has been used to predict the abundances of the primordial elements. Carrying this analogy further, just as the measurements of the light element abundances have been used to constrain the composition and evolution of our universe since the time of proton-neutron freeze out, $t \sim 1 \, {\rm s}$, measurements of the dark matter abundance and the characteristics of this particle species could be used to constrain how our universe expanded and evolved since the time of dark matter's thermal freeze out. Dark matter could thus serve as a probe of our universe's very early thermal history, allowing us to constrain times as early as $t \sim 4 \times 10^{-10} \, {\rm s} \times ({\rm TeV}/m_{\chi})^2$ after the Big Bang, many orders of magnitude prior to the onset of BBN.

\section{BBN As A Probe of New Physics: A Mini-Review}\label{sec:BBN}

Historically speaking, the light element abundances provided early and critical support for the Big Bang theory over steady-state cosmological models~\cite{Gamow:1946eb,Alpher:1948ve,Alpher:1950zz,Alpher:1953zz,Hayashi:1950lqo}. In more recent decades, such measurements have served as an important probe of our universe's early thermal history. In this section, we briefly review the physics of BBN, and how measurements of the primordial element abundances have been used to constrain new physics. This review is to illustrate the similarity between the establishment of light element abundance during BBN and dark matter freeze-out. Readers that are familiar with BBN and its use as a probe of new physics can feel free to skip the remainder of this section.

Prior to BBN, neutrons and protons were maintained in chemical equilibrium through the processes $n+\nu_e \leftrightarrow p+e^{-}$ and $n+e^{+} \leftrightarrow p +\bar{\nu}_e$. In this equilibrium state, the neutron-to-baryon ratio, $X_n$, was given by
\begin{equation}
    X_n \equiv \frac{n_n}{n_b}=\frac{1}{1+e^{(m_n-m_p)/T}},
\end{equation}
where $m_n -m_p\approx 1.29$ MeV is the neutron-proton mass splitting. 
The evolution of the neutron abundance is governed by the following Boltzmann equation:
\begin{equation}
    \frac{dX_n}{dt}=-\lambda_{np}(1+e^{-(m_n-m_p)/T})(X_n-X_n^{\rm eq}),
\end{equation}
where $\lambda_{np}$ is the neutron-proton conversion rate. In the absence of new physics, the neutron-to-baryon ratio freezes out and becomes nearly constant (up to the effects of neutron decay) at a temperature of $T\sim 1 \, {\rm MeV}$, after which $X_n\approx 0.157$. The production of deuterium and other light elements starts around $T\sim 0.07$ MeV, by which time neutron decay has reduced this ratio to $X_{n,\rm BBN} = 0.157 \, \, e^{-t_{\rm BBN}/\tau_{n}}\approx 0.122$, where $\tau_n \approx 880.2 \, {\rm s}$.
%
%
As the deuteron abundance increased, other nuclear reactions became possible, ultimately causing most of the neutrons to become bound within $\rm \prescript{4}{} He$ nuclei, which has the highest binding energy per nucleon among all isotopes lighter than carbon. As a consequence, the resulting helium mass fraction, $Y_p$, is approximately equal to half the neutron abundance at the onset of primordial nucleosynthesis:
\begin{equation}
\label{eq:Yp}
    Y_p \equiv \frac{4n_{\rm He}}{n_b}\approx 2 X_n \approx 0.245.
\end{equation}

\subsection{Exotic Matter or Radiation}

The measured abundances of primordial helium and deuterium are in good agreement with the predictions of standard BBN. Since these predicted abundances are quite sensitive to the rate of Hubble expansion (and thus to the total energy density) at the time of proton-neutron freeze out, these measurements allow us to place significant constraints on the densities of matter and other energy that were present $\sim 1 \,{\rm s}$ after the Big Bang. 

According to the first Friedmann equation, the rate of Hubble expansion is proportional to the square root of the total energy density (in the absence of significant curvature). As a result, the presence of any exotic matter or radiation will increase the expansion rate, causing the neutron-to-baryon ratio to freeze out of equilibrium earlier, when the equilibrium value of this ratio was higher. Although increasing the rate of expansion also suppresses the effectiveness of subsequent fusion processes, this turns out to be a secondary effect. 

In the case of exotic radiation, it is conventional to express the contribution to the energy density in terms of the effective number of neutrino species, $\Delta N_{\rm eff}=N_{\rm eff}-3.046$, and the energy density in photons:
\begin{align}
\rho_{\rm exotic} = \Delta N_{\rm eff} \times \frac{7}{8} \bigg(\frac{4}{11}\bigg)^{4/3} \rho_{\gamma}.
\end{align}
The ultimate impact of modifying the energy density in exotic radiation is to change the helium mass fraction by $\Delta Y_p \approx 0.013 \times \Delta N_{\rm eff}$, and the relative deuterium abundance by $\Delta (^2{\rm H}/{\rm H})_p \approx (0.325 \times 10^{-5}) \times \Delta N_{\rm eff}$. After performing a fit to the measured helium and deuterium abundances, allowing both $N_{\rm eff}$ and the baryon abundance to float, one obtains a constraint on $2.3 < N_{\rm eff} < 3.4$ (at the 95\% confidence level), consistent with the Standard Model prediction of 3.046~\cite{Cyburt:2015mya}. In other words, the density of exotic radiation at the time of proton-neutron freeze out was no more than $\sim 10\%$ of that in photons. Similar constraints have been derived on the density of exotic matter at the time of proton-neutron freeze out (for an example, see Ref.~\cite{Keith:2020jww}).

\subsection{Varying Masses and Couplings}

If any of the masses or couplings that are relevant to the processes of BBN change with time or environment, this could impact the resulting primordial element abundances. In particular, the temperature of proton-neutron freeze out is highly sensitive to the neutron-proton mass splitting, $m_n-m_p$. Furthermore, the rates for the various relevant fusion processes can be affected by the binding energies of deuterium and the other light elements. These considerations allow us to place a tight constraint of $|\Delta (m_n-m_p)|/(m_n-m_p) \lesssim 0.02$, as evaluated at the time of proton-neutron freeze out. Similarly, this information can be used to significantly constrain any change in the deuteron's binding energy, $|\Delta E_d|/E_d\lesssim 0.1$~\cite{Nollett:2002da,Dmitriev:2003qq,Coc:2006sx,Dent:2007zu}. 

 
\subsection{Energy Injection}

In many new physics scenarios, energetic particles can be injected into the universe during or after the era of BBN, potentially impacting the primordial element abundances. For example, exotic particles could have been decaying during our universe’s first minutes, creating high-energy photons and other particles that could be capable of disassociating nuclei.

An energetic photon can disassociate a nucleus in the early universe only if its energy exceeds the binding energy of its target and it is able to reach its target before being absorbed through the process of pair production. Consider, for example, a photon with just enough energy to disassociate a helium-4 nucleus, $E_{\gamma} = 28.3 \, {\rm MeV}$. At temperatures greater than a fraction of a keV, such photons will be efficiently absorbed through pair production, strongly limiting their ability to disassociate nuclei~\cite{Kawasaki:1994af}. At higher temperatures, nuclei can still be broken up by energetic pions, but not efficiently by photons.

Once the temperature has dropped enough to make photodisassociation potentially effective, any energetic photons that are injected into the bath can break up helium-4 nuclei. This process both reduces the abundance of helium, and also generates additional deuterium. Roughly speaking, such scenarios can be ruled out if the number of helium-disassociating photons exceeds the uncertainty on the measured number of primordial deuterons. 

As an example, consider an exotic particle species with a lifetime of $\sim 10^8 \, {\rm s}$ which produces one 30 MeV photon in each of its decays. If we assume that each of these photons produces one deuteron through the disassociation of a helium nucleus, then this would lead to a change in the deuteron abundance that is given by $\Delta(^2{\rm H}/{\rm H}) \sim  (\rho_X /\rho_b )(m_p /m_X)$, where $\rho_X$ and $m_X$ are the density and mass of the decaying particle. If $\rho_X /\rho_b \gsim 3 \times 10^{-6} \times (m_X /100\,{\rm MeV})$, this would produce enough additional deuterium to increase its primordial abundance beyond the measured range, allowing us to place strong constraints on particle species that were decaying or otherwise injecting energy into the universe during or shortly after the era of BBN~\cite{Keith:2020jww,Kawasaki:2017bqm,Forestell:2018txr,Hufnagel:2017dgo,Hufnagel:2018bjp,Poulin:2015opa,Cyburt:2002uv}.

\subsection{Constraints on Dark Matter}

If the dark matter was in equilibrium with the Standard Model bath in the early universe, the abundance of that species will freeze out when the rate of Hubble expansion exceeds that of dark matter annihilation. For a dark matter candidate that freezes out with an abundance that is equal to the measured density of dark matter, this freeze out occurs at a temperature approximately given by $T_f \sim m_{\chi}/20$. If the dark matter freezes out of equilibrium after proton-neutron converting processes do, the energy density of the dark matter and its annihilation products will significantly contribute to the total energy density, increasing the rate of Hubble expansion and the resulting abundance of neutrons, ultimately leading to larger abundances of primordial helium and deuterium. These considerations allow us to rule out thermal relics as dark matter candidates if they are lighter than $m_{\chi} \sim 3-10 \, {\rm MeV}$, depending on the number of their internal degrees-of-freedom, and on whether they annihilate into electrons/photons or into neutrinos and other invisible radiation~\cite{Depta:2019lbe,Jedamzik:2009uy,Boehm:2013jpa,Nollett:2014lwa}.

\section{The Abundance of a Thermal Relic}

Before turning our attention to examples of new physics scenarios that we hope to constrain using measurements of the dark matter and its abundance, we will briefly review the process of dark matter freeze out and the calculation of dark matter's thermal relic abundance.

For concreteness, we will consider here a stable dark matter candidate, $\chi$, of mass $m_{\chi}$, which we take to be a Majorana fermion. The evolution of the number density of this species in the early universe is governed by the following Boltzmann equation:
\begin{equation}
    \frac{dn_{\chi}}{dt}+3Hn_{\chi}=\langle \sigma v\rangle (n_{\chi, {\rm eq}}^2-n_{\chi}^2),
\end{equation}
where $H$ is the rate of Hubble expansion and $\langle \sigma v\rangle$ is the dark matter's thermally averaged annihilation cross section. When the $\chi$ population is non-relativistic, the equilibrium abundance of these particles is given by $n_{\chi, {\rm eq}}=g_{\chi}(m_{\chi}T/2\pi)^{3/2}{\rm exp}(-m_{\chi}/T)$, where $g_{\chi}=2$ for a Majorana fermion.

The numerical solution to this equation yields a current density that is given by~\cite{Steigman:2012nb}
\begin{align}
\label{Eq:rhoDM}
\rho_{\chi} \approx \rho_{\rm DM} \times \bigg(\frac{2.2 \times 10^{-26} \, {\rm cm}^3/{\rm s}}{\langle \sigma v \rangle}\bigg) \, \bigg(\frac{80}{g_{\star}(T_f)}\bigg)^{1/2} \, \bigg(\frac{x_f}{26}\bigg),
\end{align}
where $\rho_{\rm DM}$ is the measured dark matter density, $g_{\rm star}(T_f)$ is the number of relativistic degrees-of-freedom at the temperature of freeze out, and $x_f \equiv m_{\chi}/T_f$. This latter quantity is given by
\begin{align}
x_f \approx 26 + \ln \bigg[&\bigg(\frac{2.2 \times 10^{-26} \, {\rm cm}^3/{\rm s}}{\langle \sigma v \rangle}\bigg) \, \bigg(\frac{80}{g_{\star}(T_f)}\bigg)^{1/2} \, \bigg(\frac{g_{\chi}}{2}\bigg) \nonumber \\
&\,\,\,\,\,\,\,\,\,\,\,\, \bigg(\frac{x_f}{26}\bigg)^{3/2} \, \bigg(\frac{T_f}{10 \, {\rm GeV}}\bigg)\bigg].
\label{Eq:xf}
\end{align}

In the case of non-relativistic dark matter particles, the thermally averaged annihilation cross section that appears in Eqs.~\ref{Eq:rhoDM} and~\ref{Eq:xf} can be expressed as follows:
\begin{align}
\langle \sigma v \rangle &\approx \frac{\int \sigma v \, v^2 \,  e^{-xv^2/4} \, dv}{\int v^2 \,  e^{-xv^2/4} \, dv} \\
&= \frac{x^{3/2}}{2 \pi^{1/2}} \int \sigma v \, v^2 \, e^{-xv^2/4} \, dv, \nonumber
\end{align}
where $\sigma v$ is the dark matter's (non-thermally averaged) annihilation cross section, and $x \equiv m_{\chi}/T_f$.

\section{Exotic Matter and Radiation}\label{sec:thermal_history}

\begin{figure}[t]
\centering
\includegraphics[width=8.5cm]{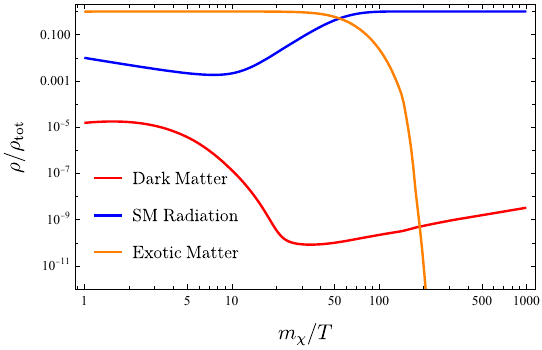}
\caption{The fraction of the total energy density that consists of Standard Model radiation, dark matter, and an exotic species of unstable matter. In this example, we have taken the exotic matter to be a scalar with 1 degree of freedom,  a mass of $m_{\Phi}=865 \, {\rm TeV}$, and a lifetime of $\Gamma_{\Phi}^{-1} = 2.2 \times 10^{-6} \, {\rm s}$. We further assume that it decoupled from the Standard Model bath at a high temperature. Once this species becomes non-relativistic, it quickly comes to dominate the energy density of the universe. It then decays, after the dark matter has frozen out of equilibrium, reheating the universe to a temperature of $T=0.3 \, {\rm GeV}$. The dark matter, in this example, is taken to be a Majorana fermion with a mass of $m_{\chi}=30 \, {\rm GeV}$ and an annihilation cross section of $\langle \sigma v \rangle = 2.2 \times 10^{-26} \, {\rm cm}^3/{\rm s}$.}
\label{fig:energy_fraction}
\end{figure}

As a first example, we will consider those scenarios in which exotic forms of matter or radiation were present during or after the era of dark matter freeze out, impacting the expansion history and the resulting thermal relic abundance of dark matter.

In the case of exotic radiation, the energy density can be expressed in terms of its contribution to the effective number of relativistic degrees-of-freedom, $g_{\star}(T)$. The total energy density is related to $g_{\star}(T)$ as follows:
\begin{align}
\rho = \frac{\pi^2 g_{\star}(T)T^4}{30},
\end{align}
where $T$ is the temperature of the photons in the bath. In kinetic and chemical equilibrium, a relativistic ($m\ll T$) fermionic (bosonic) particle species contributes 7/8 (1) to $g_{\star}$ for each of its internal degrees-of-freedom.

\begin{figure}[t]
\centering
\includegraphics[width=8.5cm]{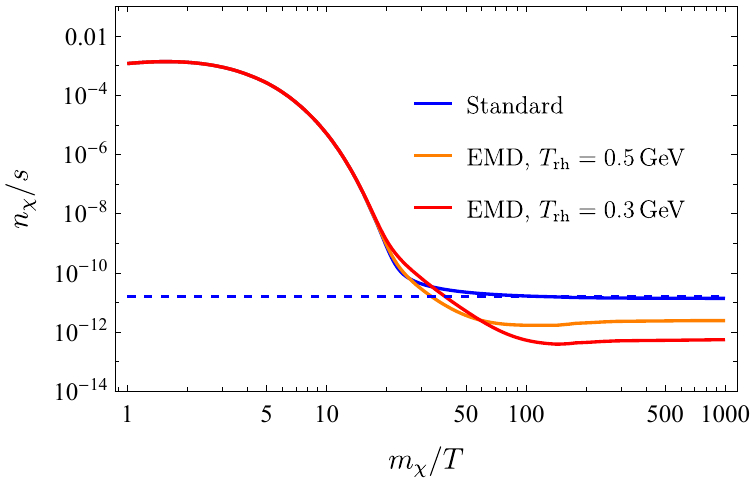}
\caption{The evolution of the dark matter number density to entropy ratio as a function of temperature for a dark matter candidate that is a Majorana fermion with a mass of $m_{\chi}=30 \, {\rm GeV}$ and an annihilation cross section of $\langle \sigma v \rangle = 2.2 \times 10^{-26} \, {\rm cm}^3/{\rm s}$. The blue curve represents the case of standard thermal freeze out, while the orange and red curves illustrate the results that are obtained in scenarios in which there was an era in the early universe that was dominated by exotic matter, resulting in the reheating of the universe to a temperature of either 0.5 or 0.3 GeV.}
\label{fig:relic_EMD}
\end{figure}

From Eq.~\ref{Eq:rhoDM}, it can be shown that (up to logarithmic corrections to $x_f$; see Eq.~\ref{Eq:xf}) the dark matter abundance scales with $g_{\star}^{-1/2}$, as evaluated at the temperature of freeze out. Thus, if we were to measure the dark matter's annihilation cross section to a precision of $p \equiv \Delta \langle \sigma v\rangle/ \langle \sigma v \rangle$, we could constrain the value of $g_{\star}(T_f)$ to lie within a range of $\Delta g_{\star}/g_{\star} \sim 2 p$. For an example in which $T_f = 1 \, {\rm GeV}$ (corresponding to $g_{\star} \approx 72$), a cross section measurement with a precision of $p=0.2$ would be sensitive to the presence of 30 (33) bosonic (fermionic) exotic relativistic degrees-of-freedom in equilibrium with the thermal bath.

Next, we will consider the case of an exotic species of matter, $\Phi$, that was present in the early universe. If such particles are long-lived and were not in equilibrium in the early universe, their abundance will evolve as $\rho_{\Phi} \propto a^{-3}$, while the energy density in radiation drops as $\rho_{\rm rad} \propto a^{-4} \, g_{\star}/g^{4/3}_{\star, S}$, where $g_{\star, S}$ is the number of relativistic degrees-of-freedom in entropy. As a result of this evolution, the decoupled matter will make up an increasingly large fraction of the total energy density as the universe expands, potentially leading to an era of early matter domination. When the exotic species eventually decays, it will heat the Standard Model bath. If the temperature after those decays is greater than a few MeV, BBN will proceed as usual, making any such scenario consistent with the measurements of the light element abundances. 

To calculate the dark matter's thermal relic abundance in this class of scenarios, we will consider the following system of Boltzmann equations:
\begin{equation}
    \begin{split}
        \frac{d\rho_{\Phi}}{dt}&+3H \rho_{\Phi}=-\Gamma \rho_{\Phi},\\
        \frac{d\rho_{\rm rad}}{dt}&+4H\rho_{\rm rad}=\Gamma \rho_{\Phi},\\
        \frac{dn_{\chi}}{dt}&+3H n_{\chi}=\langle \sigma v\rangle (n_{\chi, {\rm eq}}^2-n_{\chi}^2),
    \end{split}
\end{equation}
where $\Gamma$ is the decay width of $\Phi$. Note that in calculating the Hubble rate, $H$, we must take into account the total energy density, including that of the $\Phi$ population.

\begin{figure*}[t!]
\centering
\includegraphics[width=8.5cm]{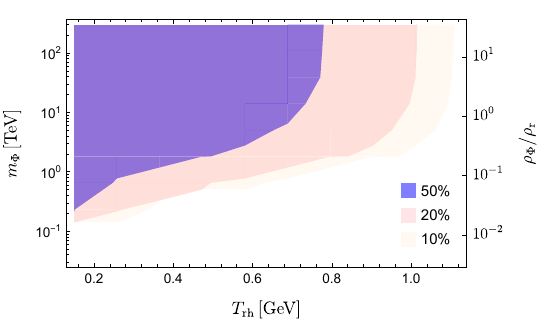}
\hspace{0.5cm}
\includegraphics[width=8.5cm]{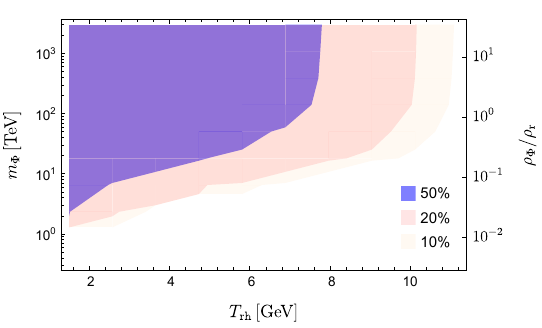}
\hspace{0.5cm}
\includegraphics[width=8.5cm]{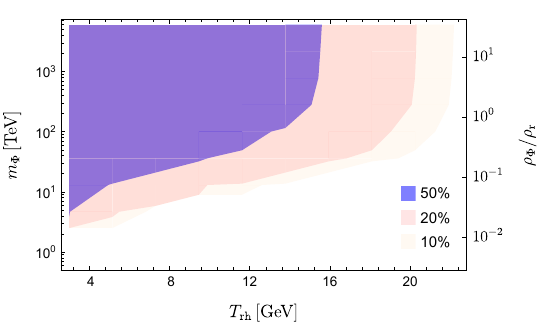}
\hspace{0.5cm}
\includegraphics[width=8.5cm]{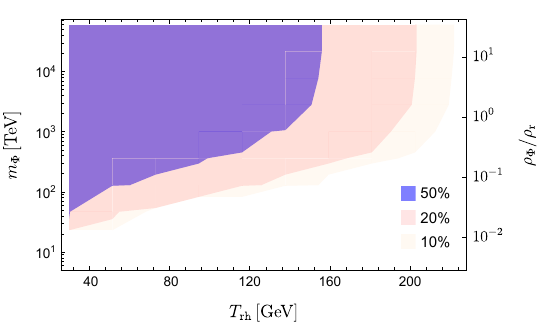}
\caption{The potential of a future measurement of the dark matter's annihilation cross section to constrain early matter dominated scenarios. The shaded regions illustrate the regions of the $m_{\Phi}-T_{\rm rh}$ plane that could be excluded by a measurement of $\langle \sigma v\rangle$ to a precision of either 50\%, 20\% and 10\%. Such a measurement could be used to rule out scenarios which feature a long matter dominated era which ends after dark matter freeze out. In the top left (right) frame, we take the mass of the dark matter particle to be $m_{\chi}=$30 GeV (300 GeV). Similarly, in the bottom left (right) frame, we take dark matter mass $m_{\chi}=$600 GeV (6000 GeV).
Along the right axis of each frame, we show the fraction of the energy density in exotic matter, as evaluated at $T=m_{\chi}$. Note that the constraints presented here can be directly rescaled to consider different reheating temperatures since the dark matter relic abundance in scenarios with early matter domination is determined by the quantities $m_{\Phi}/m_{\chi}$ and $T_{\rm rh}/T_f$.
}
\label{fig:mphi_trh}
\end{figure*}

In Fig.~\ref{fig:energy_fraction}, we show an example of an early matter dominated scenario, including a heavy scalar with $g_{\Phi}=1$ degrees of freedom, a mass of $m_{\Phi}=865 \, {\rm TeV}$, a lifetime of $\Gamma_{\Phi}^{-1} = 4.5 \times 10^{-6} \, {\rm s}$, and that decoupled from the Standard Model bath at a high temperature. Once this species becomes non-relativistic, it quickly comes to dominate the energy density of the universe. In decaying after dark matter freeze out, the exotic particles reheat the universe to a temperature of $T=0.3 \, {\rm GeV}$.

Note that so long as the exotic matter dominated the energy density prior to its decay, the reheating temperature is directly related to the decaying particle's width, $\Gamma$. This can estimated by treating the decays as instantaneous (occuring at $t=\Gamma^{-1}$), and equating the energy density of the universe before the decays,
\begin{align}
\rho_{\Phi} = \frac{M^2_{\rm Pl}}{6 \pi t^2} = \frac{M^2_{\rm Pl} \Gamma^2}{6 \pi},
\end{align}
to the energy density of the radiation bath afterwards,
\begin{align}
\rho_{\rm rad} = \frac{\pi^2 g_{\star}(T_{\rm rh}) T_{\rm rh}^4}{30},
\end{align}
yielding the following temperature of reheating:
\begin{align}
T_{\rm rh} = \bigg(\frac{5\Gamma^2 M^2_{\rm Pl}}{\pi^3 g_{\star}(T_{\rm rh})}\bigg)^{1/4}.
\end{align}

In Fig.~\ref{fig:relic_EMD}, we show an example of how dark matter freeze out could be impacted by the presence of long-lived exotic matter that was out-of-equilibrium in the early universe. In this case, the dark matter is taken to be a 30 GeV Majorana fermion with an annihilation cross section of $\langle \sigma v\rangle = 2.2 \times 10^{-26} \, {\rm cm}^3 {\rm s}^{-1}$. In the cases shown, the exotic matter impacts the resulting dark matter abundance in two distinct ways. First, the density of the exotic matter increases the total energy density at the time of freeze out, causing the dark matter to decouple earlier and increasing the resulting relic abundance. Second, the decays of the exotic matter increase the energy density of Standard Model radiation, diluting the abundance of dark matter. This will be the case so long as the thermal relic in question freezes out of equilibrium before the end of the early matter dominated era. For lower values of $T_{\rm rh}$, the matter dominated era lasts longer, allowing the exotic matter to make up a larger fraction of the total energy density before decaying, and thus leading to a greater degree of dilution and to a lower abundance of dark matter.    

In Fig.~\ref{fig:mphi_trh}, we illustrate how one could constrain scenarios featuring an early matter dominated era by measuring the dark matter's annihilation cross section to a precision of either 50\% or 20\%. In each frame, we show the regions of the $m_{\phi}-T_{\rm rh}$ plane that could be excluded by such a measurement.

\section{Ultralight scalars coupled to dark matter} \label{sec: new interactions}

Ultralight scalars have been motivated by a wide range of theoretical considerations, and include examples such as the QCD axion~\cite{Peccei:1977hh} and string theory moduli~\cite{Svrcek:2006yi,Arvanitaki:2009fg}.
Such particles can have large occupation numbers in phase space, allowing them to be described as classical waves. With small couplings to Standard Model fields, it can be challenging to detect these particles in laboratory experiments. Although ultralight scalars with universal couplings are constrained by fifth-force experiments~\cite{Olive:2007aj}, supernova cooling \cite{Raffelt:1990yz}, and black hole superradiance \cite{Arvanitaki:2010sy,Arvanitaki:2014wva}, these constraints are typically weaker than those derived from measurements of the primordial element abundances~\cite{Sibiryakov:2020eir}.

\begin{figure}[t]
\centering
\includegraphics[width=8.5cm]{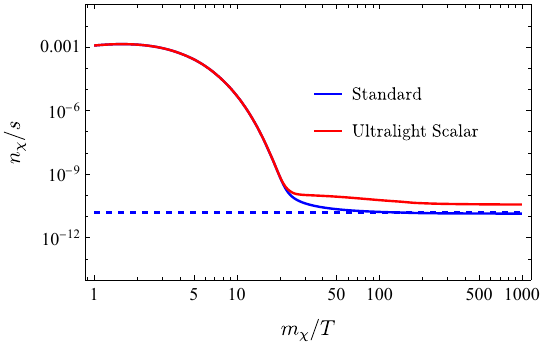}
\caption{The evolution of the dark matter number density to entropy ratio as a function of temperature for a dark matter candidate that is a Majorana fermion with a mass of $m_{\chi}=30 \, {\rm GeV}$ and an annihilation cross section of $\langle \sigma v \rangle = 2.2 \times 10^{-26} \, {\rm cm}^3/{\rm s}$. The blue curves represents the case of standard thermal freeze out, while the red curve illustrates the results that are obtained in a scenario in which the interactions of an ultralight scalar modifies the expansion rate during dark matter freeze out. In this plot, we adopt $m_{\phi}=10^{-10}$ eV and $\Lambda = 10^{13.8}\, {\rm GeV}$, corresponding to
$\langle \phi \rangle^2/\Lambda^2 \approx 0.1$. For these parameters, the interactions of the ultralight scalar increase the resulting dark matter's relic abundance by a factor of 2.7.}
\label{fig:m_Lambda1}
\end{figure}

In this section, we will focus on how an ultralight scalar coupled to dark matter could impact the expansion rate during the time of dark matter freeze out. In this context, we will consider a model that is described by the following Lagrangian:
\begin{equation}
    \mathcal{L}\supset \frac{1}{2}(\partial_{\mu} \phi)^2-\frac{1}{2}m_{\phi}^2\phi^2-\frac{\phi}{\Lambda} m_{\chi}\bar{\chi}\chi,
\end{equation}
where $\chi$ is the fermionic dark matter candidate, $\phi$ is the ultralight scalar, and $\Lambda$ is a free parameter. The ultralight scalar, $\phi$, will have a vacuum expectation value (VEV) in the early universe, $\langle \phi \rangle$, and constraints from the light element abundances require that this VEV should not be too large during the BBN era, $\langle\phi\rangle^2/\Lambda^2\lesssim 0.1$~\cite{Sibiryakov:2020eir,Stadnik:2015kia}. At earlier times, however, the value of $\langle \phi \rangle$ may have been much larger. The era of dark matter freeze out thus provides us with an powerful opportunity to probe the existence of ultralight scalars.

The evolution of $\phi$ is governed by the following differential equation:
\begin{equation}
\label{eq:eom}
    \Ddot{\phi}+3H\dot{\phi}+m_{\phi,\rm eff}^2 \, \phi=0.
\end{equation}
The effective scalar mass, $m_{\phi,\rm eff}$, receives contributions from any non-relativistic particle species that it interacts with. At early times, however, Standard Model particles disappear from the thermal bath as they become non-relativistic, and thus do not significantly contribute to the value of $m_{\phi,\rm eff}$. The dark matter, in contrast, is stable, and thus can impact $m_{\phi,\rm eff}$ more substantially:
\begin{equation}
    m_{\phi,\rm eff}^2=m_{\phi}^2+\frac{2\Theta_{\chi}}{\Lambda^2},
\end{equation}
where
\begin{align}
\Theta_{\chi}&= \rho_{\chi}-3p_{\chi}\\ 
&= 2 \frac{m_{\chi}^2T^2}{2\pi^2}\int_{m_{\chi}/T}^{\infty} dx \frac{\sqrt{x^2-m_{\chi}^2/T^2}}{e^x+1}.\nonumber
\end{align}

\begin{figure}[t]
\centering
\includegraphics[width=8.5cm]{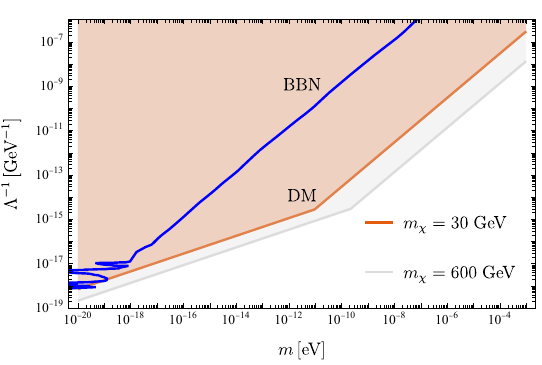}
\caption{Regions of the $m_{\phi}$, $\Lambda^{-1}$ plane in which $\langle\phi\rangle^2 / \Lambda^2 > 0.1$ at the time of proton-neutron freeze out (labeled ``BBN'')~\cite{Sibiryakov:2020eir}, or at the time of dark matter freeze out (labeled ``DM''). The region above the blue curve is excluded by BBN. The orange and gray region can be potentially probed by dark matter with masses $m_\chi=30$ GeV and $m_{\chi}=600$ GeV respectively.
We have taken the dark matter candidate to be a Majorana fermion with an annihilation cross section of $\langle \sigma v \rangle \approx 2.2 \times 10^{-26} \, {\rm cm}^3/{\rm s}$. From this comparison, it is clear that dark matter could potentially be used to probe large regions of parameter space that are inaccessible to constraints based on the primordial element abundances. A larger dark matter mass can even provide a stronger probe.
}
\label{fig:m_Lambda}
\end{figure}

From Eq.~\ref{eq:eom}, it follows that the field will start to oscillate when $m_{\phi,\rm eff}\sim H$, after which it behaves as a population of non-relativistic particles, $\rho_{\phi} \propto a^{-3}$. Prior to this transition, the value of $\phi$ remained approximately fixed at its initial value, constituting a large contribution to the overall energy density. Writing the number density and energy density of the scalars as $n_{\phi} = m_{\phi} \phi^2/2$ and $\rho_{\phi} = n_{\phi} m_{\phi, {\rm eff}} =  m_{\phi} m_{\phi, {\rm eff}} \phi^2/2$, we can express the value of $\langle\phi\rangle^2 / \Lambda^2$ at the time of dark matter freeze out as follows:
\begin{equation}
    \frac{\langle\phi\rangle^2}{\Lambda^2} \approx \frac{2 \rho_{\phi,0}}{\Lambda^2 m_{\phi} m_{\phi,\rm eff}}
    {\rm min}\left(a_{\rm fo}^{-3},a_{\rm crit}^{-3} \right),
\end{equation}
where $\rho_{\phi,0}$ is the energy density of ultralight scalars in the current universe. For concreteness, we will take $\rho_{\phi,0}$ to be 1\% of the total dark matter density. The quantities $a_{\rm fo}$ and $a_{\rm crit}$ are the scale factors evaluated at the time of dark matter freeze out and when $m_{\phi,\rm eff}=H$, respectively. 

Although the value of $\langle \phi \rangle/\Lambda$ can impact a variety of effective couplings, we will focus here on its impact on the Hubble rate through the value of $M_{\rm Pl}$. The effective Planck mass is modified through the interactions of the ultralight scalar as $M^2_{\rm Pl, \,eff} \approx M^2_{\rm Pl} \,(1-2\langle\phi\rangle^2 / \Lambda^2)$, where $M_{\rm Pl} \approx 1.22 \times 10^{19} \, {\rm GeV}$ is the standard value of the Planck mass. 

By decreasing the value of the effective Plank mass, the interactions of the ultralight scalar increase the Hubble rate, and thus cause dark matter to freeze out earlier than it would have otherwise. An example of this is shown in Fig.~\ref{fig:m_Lambda1}, for the case of $m_{\phi}=10^{-10}$ eV and $\Lambda = 10^{13.8} \, {\rm GeV}$, corresponding to $\langle\phi\rangle^2 / \Lambda^2 \approx 0.1$ at the time of dark matter freeze out. As in the previous section, we have considered here a dark matter candidate that is a Majorana fermion with a mass of $m_{\chi}=30 \, {\rm GeV}$ and an annihilation cross section of $\langle \sigma v \rangle = 2.2 \times 10^{-26} \, {\rm cm}^3/{\rm s}$. For this choice of parameters, the interactions of the ultralight scalar increase the dark matter's relic abundance by a factor of 2.7.

In Fig.~\ref{fig:m_Lambda}, we show the regions of the $m_{\phi}$, $\Lambda^{-1}$ plane in which the value of $\langle\phi\rangle^2 / \Lambda^2$ is greater than 0.1 at the time of proton-neutron freeze out~\cite{Sibiryakov:2020eir}, and at the time of dark matter freeze out. From this, it is clear that dark matter could potentially be used to probe large regions of this parameter space that is inaccessible to constraints based on BBN.

\section{Modified Gravity Scenarios} \label{sec: mod_gravity}

Modifications to general relativity have been motivated by many considerations, including those of dark energy and dark matter, as well as by challenges encountered in efforts to renormalize gravity~\cite{Stelle:1976gc}. In some classes of modified gravity theories, the effects are most significant in the early universe, leading to strong constraints from BBN~\cite{Kang:2008zi, Kusakabe:2015yaa, Asimakis:2021yct}, and perhaps one day from the freeze out of dark matter. In this section, we briefly consider a few examples of such scenarios.

\subsection{$f(G)$ gravity}

As a first example, we will consider a class of modified gravity models in which the Einstein action is modified by the quantity, $f(G)$, which is a function of the Gauss-Bonet invariant, $G$~\cite{Nojiri:2005jg}. The modified Einstein action in such models is given by 
\begin{equation}
    S=\int d^4 x \, \sqrt{-g}\left[\frac{M^2_{\rm Pl}}{128 \pi^2}R+f(G)\right],
\end{equation}
where $g$ is the determinant of the metric tensor, and $R$ is the Ricci scalar. The quantity, $G$, is given by~\cite{Nojiri:2005jg}
\begin{equation}
    G= R^2-4R_{\mu\nu}R^{\mu\nu}+R_{\mu\nu\rho\sigma}R^{\mu\nu\rho\sigma}.
\end{equation}
In FRW metric the Gauss-Bonnet invariant can be expressed as $G=24H^2(H^2+\dot{H})$.
The new $f(G)$ term leads to a modification of gravity which takes the form of a new contribution to the energy density:
\begin{equation}
\begin{split}
        \rho_{f(G)}= &\frac{1}{2}\Bigl[ -f(G)+24H^2(H^2+\dot{H})f^{\prime}(G) \\
        &-24^2H^4(2\dot{H}^2+H\ddot{H}+4H^2\dot{H})f^{\prime\prime}(G) \Bigl].
\end{split}
\end{equation}


Adopting the following form $f(G)=\alpha G^n$, one finds that this contribution to the energy density scales as follows:
\begin{equation}
    \frac{\rho_{f(G)}}{\rho_{\rm SM}}\sim \alpha(128\pi^2 G_N^2\rho_{\rm SM}^2/3)^n\rho_{\rm SM}^{-1} \propto \alpha \, T^{8n-4}.
\end{equation}
Thus, if $n>1/2$, the exotic component of the energy density will be most significant at the highest temperatures and earliest times, making dark matter freeze out a powerful probe of this class of scenarios. Taking $n=1$ as an example, we find that the dark matter abundance can be used to place a constraint of $\alpha\lesssim 10^{67}$ if we require ${\rho_{f(G)}}/{\rho_{\rm SM}}<0.1$ at $T=10$~GeV. This number may look large, but for this choice of parameters, the exotic energy density is only $\Omega_{f(G)} \sim 10^{-54}$ in the current universe.

\subsection{Running Vacuum Model}

Models in which gravitational anomalies are coupled to axion fields \cite{Basilakos:2019acj} can lead to time-dependent contributions to the energy density. The exotic energy density in such ``running-vacuum'' models can be expressed as \cite{Shapiro:2009dh,Geng:2017apd,Mavromatos:2021urx}
%
%
%
\begin{equation}\label{eq:rvm}
    \rho_{\rm RVM} = a +b H^2 +c H^4+...
\end{equation}
where the $a$ term corresponds to the cosmological constant, and the $b$ and $c$ terms represent deviations from the $\Lambda{\rm CDM}$ paradigm. In the early universe, the $c H^4$ term will generally be most significant, such that
\begin{equation}
    \frac{\rho_{\rm RVM}}{\rho_{\rm SM}}= \frac{8\pi G_N}{3}c H^2 =g_{\star}(T)\frac{32\pi^4}{135} c \;G_N^2 T^4 .
\end{equation}
$G_N$ represents Newton's constant, which should not be confused with the Gauss-Bonnet invariant $G$.
Once again, this implies that probes of our universe's very early thermal history could be used to place strong constraints on such scenarios. By requiring $\rho_{\rm RVM}/\rho_{\rm SM}<0.1$ at $T =10$~GeV, we can constrain $c\lesssim 10^{68}$ in this model.


In addition to those we have considered here, there are many other well-motivated theories of modified gravity which share these general features, including $f(P)$ gravity~\cite{Lovelock:1971yv}, and Gauss-Bonnet-Dilaton Gravity~\cite{Kanti:1995vq}. In scenarios in which the effects of new physics are most pronounced at high temperatures, we could expect dark matter freeze out to potentially provide very strong constraints.


\section{Discussion and Summary}

In this paper, we have imagined a time in the future when we have not only detected the particle that makes up the dark matter of our universe, but have measured its mass and annihilation cross section with a reasonably high degree of precision. Combining this information with the measured dark matter abundance, one could probe the expansion rate of our universe at the time of dark matter's decoupling. This is in close analogy to how the light element abundances have been used to probe the era of of Big Bang nucleosynthsis, as early as $t \sim 1 \, {\rm s}$ after the Big Bang. For dark matter in the form of a generic thermal relic, freeze out occurs at a temperature of $T_f \sim m_{\chi}/20$, corresponding to $t \sim 4 \times 10^{-10} \, {\rm s} \times ({\rm TeV}/m_{\chi})^2$. This process could thus potentially be used to probe the thermal history of our universe at much earlier times than BBN.

One could reasonably ask how realistic it is that we could measure the dark matter's annihilation cross section with enough precision to meaningfully probe the thermal history of the early universe. As an illustrative example, consider a scenario motivated by the observed Galactic Center Gamma-Ray Excess~\cite{Goodenough:2009gk,Hooper:2010mq,Hooper:2011ti,Daylan:2014rsa}. This signal is consistent with arising from annihilating dark matter particles with a mass of $m_{\chi}\sim 40-50 \, {\rm GeV}$ and an annihilation cross section consistent with that expected of a thermal relic, $\langle \sigma v \rangle \sim (1-2) \times 10^{-26}\, {\rm cm}^3/{\rm s}$~\cite{Cholis:2021rpp}. In the spirit of this study, we consider a time when a large space-based gamma-ray telescope (such as the proposed Advanced Particle-astrophysics Telescope~\cite{APT:2021lhj}) has confirmed a dark matter interpretation of of the Galactic Center Gamma-Ray Excess by measuring the corresponding signal from several dwarf spherioidal galaxies. Such a measurement would ultimately be limited to our ability to determine the $J$-factors of these dwarf galaxies, the best of which are currently measured to a precision of around 30\% (see, for example, Ref.~\cite{Pace:2018tin}). By combining the observations of several such dwarf galaxies, it seems plausible that we could determine dark matter's low-velocity annihilation cross section to a precision on the order of tens of percents.

The finite temperature at the time of dark matter freeze out somewhat complicates this story, as the annihilation cross section as evaluated at $T_f$ will be different in some models than it is in the modern, low temperature universe. For example, many dark matter models feature both $s$-wave and $p$-wave contributions to the annihilation cross section, allowing us to write $\langle \sigma v \rangle \approx a + b v^2.$ If the $p$-wave term is not negligible at the time of dark matter freeze out, the relic abundance we infer from the $s$-wave term alone will not match the measured abundance. Fortunately, the results of direct detection experiments could be helpful in clarifying such a situation. From the combination of the information that could be provided by indirect and detection experiments, it is plausible that the spin and Lorentz structure of the dark matter and its interactions could be reasonably established. Among dark matter candidates that are capable of generating the Galactic Center excess, for example, one could consider a scalar Higgs portal model~\cite{DiMauro:2023tho,GAMBIT:2017gge}, or dark matter that is a fermion with mixed vector and axial-vector couplings~\cite{Berlin:2014tja}. While these models would produce potentially indistinguishable gamma-ray signals, they would lead to very different rates at future direct detection experiments, enabling us to identify the correct underlying model and to correctly infer the $p$-wave contribution to the dark matter's annihilation cross section at the time of its freeze out.

Once these characteristics of the particles that make up the dark matter have been established, a vast range of new physics scenarios could be effectively constrained and studied. In this paper, we have considered scenarios featuring exotic forms of radiation or matter, models with a ultralight scalar, and several theories of modified gravity, each of which have the potential to be far more powerfully probed with dark matter than by other means, including BBN. This list is by no means intended to be inclusive, but rather reflects the tip of the iceberg of what we could learn from such a series of measurements. For decades, our measurements of the light element abundances have given us confidence that our universe was radiation dominated and otherwise adhered to the predictions of the $\Lambda$CDM model since $\sim 1 \,{\rm s}$ after the Big Bang. Prior to this time, we have no empirical information with which to ground any such claims. In a future in which dark matter's nature has been adequately measured, we could extend this kind of understanding back in time by many orders of magnitude, mapping out our universe's thermal history as early $\sim 4 \times 10^{-10} \, \times \, ({\rm TeV}/m_{\chi})^2 \, {\rm s}$ after the Big Bang.

\section*{Acknowledgements}
We would like to thank Gordan Krnjaic, Lian-Tao Wang, and Xucheng Gan for helpful discussions.
DH and HX are supported by Fermi Research Alliance, LLC under Contract DE-AC02-07CH11359 with the U.S. Department of Energy.
\bibliographystyle{apsrev4-2}
\bibliography{wimp_cos}
\end{document}